 \documentclass[aps,pra,superscriptaddress,amsmath,amssymb,preprintnumbers,twocolumn,floatfix,showpacs,showkeys,10pt]{revtex4-1}
 \usepackage{amssymb} \usepackage{epsfig}
 \begin{document}
  \title{Time-evolution of tripartite quantum discord and entanglement  under local and non-local random telegraph noise }
   \author{Fabrizio~Buscemi }
 \email{fabrizio.buscemi@unimore.it} \affiliation{ARCES, Alma
   Mater Studiorum, Universit\`{a} di Bologna, Via Toffano 2/2, I-40125 
   Bologna, Italy}

  \author{Paolo~Bordone} \affiliation{ Dipartimento di
   Scienze Fisiche, Informatiche e Matematiche, Universit\`{a} di Modena e Reggio Emilia  and \\ Centro S3, CNR-Istituto di Nanoscienze,  Via
   Campi 213/A, I-41125, Modena, 
   Italy}

 \begin{abstract} 
Few studies explored  the dynamics of non-classical correlations  besides entanglement
in open multipartite quantum systems. Here,
 we address the time-evolution of quantum discord and entanglement
in a model  of three non-interacting qubits subject to  a classical random telegraph noise  in common
and separated environments.  Two initial entangled states of the system are examined,
namely the GHZ- and W-type states.   The dynamics of quantum correlations results  to be
strongly affected by the  input configuration of the qubits, 
the type of the system-environment interaction, and the
memory properties of  the environmental noise. 
When the  qubits are non-locally coupled to the random telegraph noise,
the GHZ-type states partially preserve, at long times, both  discord
and entanglement,
regardless the correlation time of the environmental noise.  The survived entangled states 
turn  out to be also detectable by means of suitable entanglement witnesses. 
On the other hand, in the same conditions,  the 
decohering effects suppress all the quantum correlation of the W-type states
which are thus less robust than the GHZ-type ones. The long-time survival 
of tripartite discord and entanglement opens interesting perspectives 
in the use of multipartite entangled states for practical applications 
in quantum information science.
%

.

 \end{abstract}

 \maketitle
\section{Introduction}
Entanglement is a peculiar feature of quantum mechanics
which constitutes  a  valuable resource for a number of features of
quantum information processing, such as quantum communication,
cryptography, and  computation~\cite{Nielsen}. However, the major limitation
of the use of entanglement in practical applications is due
to the unavoidable interaction of the real quantum systems
with their surroundings, resulting in decoherence processes,
and, as a consequence, in degradation of the entangled
states~\cite{Joos}. Therefore, many research efforts focused on
the investigation  of the decoherence processes
and of the dynamics of entanglement in different quantum systems  ranging
from quantum optics~\cite{Yu,Bellomo2008,Maniscalco,An} to nanophysics~\citep{BuscemiSAW,*Buscemi2010,*Buscemi2011}. The decohering effects  of
the environments with different memory properties
have widely been studied in bipartite systems
where entanglement can exhibit  peculiar features 
including  sudden death and transitions, revivals, and trapping~\cite{Yu,Bellomo2007,Bellomo2008,Maniscalco}.
On the other hand, only recently the  time evolution
of the entanglement has begun to be analyzed  in multipartite open quantum systems, since in such systems
the  characterization and quantification
of the quantum correlations still represent a hard task. Specifically, 
different approaches have been used~\cite{Casagrande, Altintas, Zhong-Xiao, Anza, An, Siomau,  Siomau2, Ma, Ann,Liu,Weinstein2009,Weinstein2010}.
Some of them
involve the generalization of entanglement measures ranging from negativity~\citep{Casagrande, *Altintas, *Zhong-Xiao, *Anza}
to concurrence~\cite{An, Siomau,*Siomau2}
or Bell nonlocality disequalities~\citep{Ma, *Ann,*Liu}  commonly adopted in bipartite systems.
Multipartite entanglement has also  been investigated
by means of suitable entanglement witnesses (EWs), namely observables that can detect
the presence of the entanglement itself~\cite{Acin,Weinstein2009,Weinstein2010}.

Apart from entanglement,  quantum states can exhibit other correlations 
not present in classical systems. In particular, in the last years
quantum discord (QD) has   received  an increasing  attention as a mean to estimate the quantum
correlations encoded in a system and potentially usable for applications
in quantum informations processing~\cite{Ollivier,Henderson, Fanchini, Mazzola,Fanchini2010}.
Indeed, it was shown theoretically~\cite{Datta},
and later experimentally~\cite{Lanyon}, that bipartite separable mixed states
with nonzero discord may provide computational
speedup compared to classical states in some
quantum computation protocols. Under
suitable conditions  quantum discord is also found to be more robust than entanglement in  noisy environments~\cite{Fanchini}.
While the characterization of quantum discord in terms of the discrepancy between two quantum analogues of the classical mutual information 
has well been grasped and widely used in bipartite systems, its extensions in multipartite systems 
is still discussed and tackled with different approaches~\cite{Kasz, Rulli,Chaka,Grimsmo, Giorgi,Zhao}. Recently, Giorgi~\emph{et al.}~\cite{Giorgi} introduced
a measure of \emph{genuine} total, classical, and quantum correlations relying on the use of
relative entropy to quantify the distance between two density matrices.  Like other 
proposed measures of correlations in quantum systems~\cite{Rulli,Chaka, Zhao}, the approach of Ref.~\cite{Giorgi}
involves difficult extremization procedures over operators or states which makes the calculations very hard.
This justifies the scarse number of works exploring the time evolution of quantum correlations
in multipartite open systems.

In this paper, we investigate the dynamics of the quantum correlations
in a physical model  consisting of three  qubits,  not interacting among each other, and
subject to a classical random telegraph noise (RTN).  Two different initial
configurations are examined, namely  GHZ and W  Werner type-states.
The influence of the classical environmental noise  on the  system is described by means of a stochastic Hamiltonian
with a  coupling term mimicking a random telegraph signal. The dynamics of the two qubits
is evaluated by averaging the time-evolved states over the noise.
Even if in similar  systems of two qubits
the time evolution of quantum correlations has already been investigated~\cite{Zhou,Zhou2011,Mottototen, BordoneFNL,*BenedettiAr}, the analysis of the tripartite regime seems quite promising since here the correlations derive from a more complex scenario, 
and, as a consequence,  more  peculiar phenomena are expected.

The use of a stochastic Hamiltonian allows us to
evaluate   the  dynamics of  qubits 
in both  separated  common environments in a straightforward way. While in the former case each subsystem is locally coupled
to a source of RTN, in the latter a non-local interaction between the entire system and the environment occurs,
being the qubits collectively coupled to the same source  of RTN.
This gives us the chance to  analyze 
how the local or non-local nature of the qubits-environment coupling
and the memory properties of the environment itself  affect the dynamics 
of quantum correlation and its possible sudden death, revival, or trapping.
 From this point of view,
 the choice of a classical environment
does not represent  a strict constraint  of the model, since a quantum environment
not affected by the system or influenced in a way that does not result in back-action,
can be mimicked by a classical noise~\cite{Saira} that was already shown to be able to lead, in bipartite systems,
to sudden death and revival of entanglement and discord~\cite{Zhou,BenedettiAr,LoFranco2012}.

Here, we focus on the quantum correlations in the form of  tripartite entanglement and  quantum discord.
Being  the characterization of tripartite mixed entangled states  
still debated,  two different estimators are adopted to evaluate
 analytically  the entanglement in agreement with what was done in other works~\cite{Weinstein2009,Weinstein2010}. One estimator is the tripartite negativity~\cite{Sabin}, which
was shown to play a key role in quantum information protocols since states exhibiting
nonzero tripartite negativity are distillable to GHZ states. The other estimator  relies on the
concept of \emph{detectable} entanglement and uses suitable EWs~\cite{Acin}.
The quantum discord is  numerically quantified by using the  approach developed in Ref.~\cite{Giorgi}.
Unlike the tripartite negativity and the expectation value of the EWs,  in this case we need to use 
 numerical techniques apt to optimize, over positive operator-valued measures (POVM's)  and subsystems, the  conditional entropies appearing in 
the definition of discord. The comparison between the time evolution 
of these estimators for two different initial configurations of the qubits  can represent
a valid guideline to shed some light on  the discrepancies and the analogies  between the two forms of correlation
in our model.

The paper is organized as follows: in Sec.~\ref{criteria}, we
introduce the estimators used to quantify  tripartite quantum discord, 
 entanglement, and \emph{detectable} entanglement. In Sec.~\ref{model},
 we illustrate the physical model consisting of three
 qubits subject to a classical RTN within common
 or different environments.  Sec.~\ref{results} reports
 the time evolution of quantum correlations for
 two initial configurations of the qubits in the case of
 local and non-local system-environment interaction.
 Finally, conclusions  and discussions are summarized in Sec.~\ref{conclusions}.

 \section{Quantum correlation measures in tripartite systems}\label{criteria}
%
In this section, we briefly expose the correlations  measures adopted in our work to quantify
and characterize the  tripartite quantum discord and  entanglement.

 \subsection{Tripartite Quantum Discord}
Here, we illustrate the criterion adopted to quantify 
 the \emph{genuine} total, classical, and quantum correlations in a tripartite system.
 By  following the approach used in Ref.~\cite{Giorgi},  the genuine total  correlations $\mathcal{T}^{(3)}(\rho)$,
  of a mixed state $\rho$ can be expressed as:
 \begin{equation}  \label {trho3}
 \mathcal{T}^{(3)}(\rho)= \mathcal{T}(\rho)- \mathcal{T}^{(2)}(\rho),
 \end{equation}
 where $\mathcal{T}(\rho)$ is  the  quantum extension of the Shannon classical mutual information
 \begin{equation}  \label {trho}
 \mathcal{T}(\rho)=S(\rho_A)+S(\rho_B)+S(\rho_C)-S(\rho)
 \end{equation}
 with $S(\rho_K)$=$-\textrm{Tr}\big(\rho_K\ln_2{\rho_K}\big)$ indicating the the von Neumann entropy
 of $\rho_K$.
$\mathcal{T}^{(2)}(\rho)$ is given by the maximum  of the total bipartite  correlations among two 
qubits  of the system, namely
 \begin{equation}  \label {trho2}
 \mathcal{T}^{(2)}(\rho)=\max{\left[\mathcal{I}^{(2)}(\rho_{A,B}), \mathcal{I}^{(2)}(\rho_{A,C}),\mathcal{I}^{(2)}(\rho_{B,C})\right]}
 \end{equation}
with $\mathcal{I}^{(2)}(\rho_{I,J})$=$S(\rho_{I})+S(\rho_{J})-S(\rho_{IJ})$ denoting
the bipartite quantum mutual information of the subsystem consisting of the two qubits $I$ and $J$.
By inserting  Eqs.~(\ref{trho}) and  (\ref{trho2}) in (\ref{trho3}), one finds
 \begin{eqnarray}  
 \lefteqn{\mathcal{T}^{(3)}(\rho)=} \nonumber \\  & &\min{\left[\mathcal{I}^{(2)}(\rho_{A,BC}), \mathcal{I}^{(2)}(\rho_{B,AC}),\mathcal{I}^{(2)}(\rho_{C,AB})\right]},
 \end{eqnarray}
that is the total correlations of the state $\rho$ are the lowest bipartite mutual information between  the one-qubit part and the two-qubit part.
As indicated in Ref.~\cite{Giorgi}, $\mathcal{T}^{(3)}(\rho)$ estimates
the amount of \emph{genuine} tripartite correlations that cannot be accounted for considering any of the possible
subsystems.

Analogously with $T^{(3)}(\rho)$ , also the \emph{genuine} classical correlations $\mathcal{J}^{(3)}(\rho)$
can be defined in terms of the difference between the total classical ones $\mathcal{J}(\rho)$
and the maximum among bipartite correlations $\mathcal{J}^{(2)}(\rho)$:
\begin{equation}
\mathcal{J}^{(3)}(\rho)= \mathcal{J}(\rho)- \mathcal{J}^{(2)}(\rho).
 \end{equation} Here
\begin{eqnarray}  
\lefteqn{ \mathcal{J}(\rho)=}\nonumber \\ & &\max_{I,J,K}{\left[ S(\rho_I)-S(\rho_{I|J})+S(\rho_K)-S(\rho_{K|IJ})\right]},
   \end{eqnarray}
denotes the maximum over the six possible
quantum extensions of the conditional entropy 
 which, in turn,
is maximized on  the complete measurement on two subparties
where
\begin{widetext}
\begin{eqnarray}  \label{ces}
& &S(\rho_{I|J})= \min_{\{E^m_J\}}\sum_m p_m S(\rho_{I|E_m^J}) \quad \textrm{with}\quad p_m=\textrm{Tr}_{IJ}\left(E_m^J \rho_{IJ}\right),
\rho_{I|E_m^J}=\frac{\textrm{Tr}_{J}E_m^J \rho_{IJ}}{p_m}   \\
& &S(\rho_{K|IJ})= \min_{\{E^I_l,E^J_m\}}\sum_{l,m} p_{lm} S(\rho_{K|E_m^JE_l^I}) \quad \textrm{with}\quad p_{lm}=\textrm{Tr}_{IJK}\left(E_l^IE_m^J \rho\right),
\rho_{K|E_m^JE_l^I}=\frac{\textrm{Tr}_{IJ}E_l^I E_m^J\rho}{p_{lm}}.  \nonumber 
\end {eqnarray}
\end{widetext}
 In the above expressions, $\{E^I_l\}$, and $\{E^J_m\}$ indicate the POVM's
 acting on the qubits $I$ and $J$, respectively. On the other hand, $\mathcal{J}^{(2)}(\rho)$ corresponds to  the maximum  of the classical bipartite  correlations among two 
qubits  of the system
 \begin{eqnarray}  \label {jrho2}
\lefteqn{ \mathcal{J}^{(2)}(\rho)=} \nonumber \\ & &\max{\left[\mathcal{J}^{(2)}(\rho_{A,B}), \mathcal{J}^{(2)}(\rho_{A,C}),\mathcal{J}^{(2)}(\rho_{B,C})\right]},
 \end{eqnarray}
 where $\mathcal{J}^{(2)}(\rho_{I,J})$=$\max\left[\mathcal{J}^{(2)}_{I:J}(\rho_{I,J}),\mathcal{J}^{(2)}_{J:I}(\rho_{I,J})\right]$
with $\mathcal{J}^{(2)}_{I:J}(\rho_{I,J})$=$S(\rho_I)- S(\rho_{I|J})$.

In agreement with the usual definition
for two-qubit systems~\cite{Ollivier,Henderson}, the tripartite quantum discord $\mathcal{D}^{(3)}(\rho)$ can be expressed as
\begin{equation} \label{tqd}
\mathcal{D}^{(3)}(\rho)= \mathcal{T}^{(3)}(\rho)- \mathcal{J}^{(3)}(\rho).
 \end{equation}

 Given the optimization procedures over the indices 
 and  POVMs appearing in the formal expression
 of $ \mathcal{T}^{(3)}(\rho)$ and  $\mathcal{J}^{(3)}(\rho)$,  the evaluation of
 Eq.~(\ref{tqd}) is not immediate. For symmetrical states,
 namely those systems whose state  is invariant under the permutations of the three parties,  it can be shown
 that the tripartite total, classical, and quantum correlations can be evaluated by
 means of simple expressions. After a straightforward calculation, one finds 
  \begin{eqnarray}  \label{stqd}
 \mathcal{T}^{(3)}(\rho)&=&S(\rho_{C})+S(\rho_{AB})-S(\rho) \nonumber  \\
 \mathcal{J}^{(3)}(\rho)&=&S(\rho_{C})-S(\rho_{C|AB}) \nonumber \\
  \mathcal{D}^{(3)}(\rho)&=&S(\rho_{AB})+S(\rho_{C|AB})- S(\rho).
 \end{eqnarray}
These are formally equivalent to the forms reported in Ref.~\cite{Zhao}
but here, in agreement with the approach developed by Giorgi~\emph{et al.}~\cite{Giorgi}
 the von Neumann entropy of the measurement-based conditional density operator $\rho_{C|AB}$ is minimized over the set  of two-qubit POVM's  factorizable in terms of single-qubit POVM's $E_m^{A}\otimes E_n^{B}$. The use of the latter  as optimal POVM in the definition 
of classical tripartite correlations has been discussed~\cite{Zhao}. In our numerical approach
the von Neumann entropy  of  the measurement-based conditional density operator given in Eqs.~(\ref{ces})  
has been estimated by using  projective measurements described by  the complete set of orthogonal projectors 
$\Pi^{AB}_{mn}(\theta_A,\theta_B, \Phi_A,\Phi_B)$=$\sum_{m,n}|\widetilde{m}_A \widetilde{n}_B\rangle \langle \widetilde{n}_B \widetilde{m}_A|$
(with $m,n=0,1$) defined by the orthogonal states
\begin{eqnarray}
| \widetilde{0}_{A(B)}\rangle&=& \cos{\theta_{A(B)} }|0\rangle +e^{i\Phi_{A(B)}} \sin{\theta_{A(B)}}|1\rangle \nonumber \\
| \widetilde{1}_{A(B)}\rangle&=& e^{-i\Phi_{A(B)}}\sin{\theta_{A(B)} }|0\rangle -\cos{\theta_{A(B)}}|1\rangle.
\end{eqnarray}
The conditional entropy  $S(\rho_{C|AB})$ has been minimized over the phases $\theta_A$, $\theta_B$, $\Phi_A$, and $\Phi_B$.

 In the followings,
we will use  Eqs.~(\ref{stqd})  to evaluate the dynamics
of quantum discord  since at any time $t$ the system
investigated is invariant under the permutations
of the three qubits.

\subsection{Tripartite Negativity} In order to quantify  the three-qubit
entanglement of a mixed state $\rho$, we use the tripartite negativity $\mathcal{N}^{(3)}$
which is given by~\cite{Sabin}:
\begin{equation}
\mathcal{N}^{(3)}(\rho)=\sqrt[3]{\mathcal{N}_{A-BC}\mathcal{N}_{B-AC}\mathcal{N}_{C-AB}}.
\end{equation}
It is the geometric mean of the bipartite negativities $\mathcal{N}_{I-JK}$
(where $I$=$A,B,C$ and $JK$=$BC, AC, AB$), which, in turn, are defined
as $\mathcal{N}_{I-JK}$=$\sum_i \left| a_i(\rho^{T_I}) \right |  -1$,
with $a_i(\rho^{T_I})$ the eigenvalues of  the partial  transpose $\rho^{T_I}$
of the total density matrix with respect to the subsystem $I$. For  symmetrical  tripartite quantum systems,
$\mathcal{N}^{(3)}$ reduces to the bipartite negativity of  any  bipartition of the system.
Recently, $\mathcal{N}^{(3)}$ has also been used to estimate the entanglement
of  tripartite bosonic and fermionic systems~\cite{Buscemi2011b}.

Even if the positivity of the tripartite negativity ensures that the state of the system under
investigation is not separable and distillable to a GHZ state, $\mathcal{N}^{(3)}$  cannot  be used to
classify and  fully characterize the entanglement of general mixed tripartite states~\cite{Sabin}
Apart from pure states,  null tripartite negativity could indeed  not imply the absence of entanglement.

\subsection{Entanglement Witnesses}
 Multipartite entanglement can also be investigated
 by means of entanglement witnesses (EWs),  namely suitable
 observables that, at least in principle, can be experimentally 
 implemented  to detect the presence of  entanglement~\cite{Acin,Weinstein2009,Weinstein2010}.
 Following the classification
 of Ref.~\cite{Acin},  four classes of three-qubit mixed states can be defined: separable
 states (S), biseparable states (B), W states, and GHZ states. Note that each class
 includes within it the previous classes as special cases.   EWs are suitable  observables
 whose  expectation value is zero  or positive 
 for all the states belonging to a specific class and negative for at least one state
 of  higher class, namely a more inclusive class.  Thus the use of EWs allows one to determine which class
 a given state belongs. 
 

 Specifically, 
 in this work we will use  those EWs  that permit to identify
 whether a state is in the $W$-$B$ class, namely a state with true tripartite
 entanglement either of the GHZ-type or W-type and not biseparable.
 As shown in Refs.~\cite{Acin,Weinstein2009}, GZH-type states in the $W$-$B$ class can be detected by means of
 \begin{equation} \label{EW1}
 \mathcal{W}_{W2}=\frac{1}{2}I-P_{GHZ},
 \end{equation}
where   $P_{GHZ}$ is the projector onto $|GHZ\rangle$.
On the other hand, for W-type states, we will use the EW
\begin{equation}  \label{EW2}
\mathcal{W}_{W1}=\frac{2}{3}I-P_W.
 \end{equation}
 with   $P_{W}$  the projector onto $|W\rangle$. 
 Negative expectation values of the observables of the Eqs.~(\ref{EW1}) and (\ref{EW2})
 indicate the appearance   of tripartite entanglement 
 experimentally detectable in the system. However, zero or
 positive values  do not  guarantee the absence of entanglement.

 \section{The model}\label{model}
Here, we describe a model consisting of three non-interacting qubits subject  to an environmental 
classical RTN. The system-environment coupling is here analyzed in two different conditions.
In the first one, each qubit  locally interacts with its environment. In the other case,
all the qubits are coupled  with a unique common source of RTN thus mimicking 
a non-local interaction between the qubits themselves and environment. 
In both configurations, the dynamics of the system is ruled by the Hamiltonian
\begin{equation} \label{mod1}
\mathcal{H}(t)=\mathcal{H}_A(t) \otimes I_{BC}+\mathcal{H}_B(t) \otimes I_{AC}+\mathcal{H}_C(t) \otimes I_{AB},
\end{equation}
where $I_{JK}$ is the identity operator in the subspace of the two qubits $J$ and $K$ and 
$\mathcal{H}_L(t)$ denotes the single-qubit Hamiltonian~\cite{BordoneFNL,BenedettiAr}
\begin{equation} \label{mod2}
\mathcal{H}_L(t) = \varepsilon I_L + \nu \eta_L(t){\sigma_x}_L,
\end{equation}
with ${\sigma_x}_L$ indicating the Pauli matrix of the subspace of the qubit $L$. $\varepsilon$ is the qubit energy in the absence of noise (energy degeneracy is assumed), $\nu$ is the system-environment coupling constant. The RTN is here introduced by means the stochastic process denoted by $\eta_L(t)$
describing a  fluctuator  randomly flipping between the values -1 and 1 at rate $\gamma$.
The two-particle  form 
of  the model defined by Eqs.~(\ref{mod1}) and (\ref{mod2})
has already been used to evaluate the time behavior of entanglement and quantum discord~\citep{BordoneFNL,*BenedettiAr}.
Furthermore its extension to qudits case allows also to analyze the quantum walks of not-interacting particles in one-dimensional lattices~\cite{Benedetti}.

Given the random nature of the telegraph processes,  the time-dependent Hamiltonian of  Eq.~(\ref{mod1}) 
leads to a stochastic evolution of the quantum states. In order to obtain the time-evolution of the system
under the influence of RTN, we need to average over
different noise configurations~\cite{Mottototen, BordoneFNL,BenedettiAr}. Therefore, the dynamics
of the system density matrix can be written as
\begin{equation} \label{rhoev}
\rho(t)=\left\langle U\left(\{\eta\},t\right)\rho(0)U^{\dag}\left(\{\eta\},t\right)\right\rangle_{\{\eta\}}
\end{equation}
where $\rho(0)$ is the initial state of the three qubits and $U\left(\{\eta\},t\right)$=$U_A\left(\eta_A,t\right) \otimes U_B\left(\eta_B,t\right) \otimes U_C\left(\eta_C,t\right)$ indicates
the unitary time evolution operator of the system at time $t$ for a
given noise configuration $\{\eta\}$=$\{\eta_A,\eta_B,\eta_C \}$
(where $\eta_A(t) \neq \eta_B(t) \neq \eta_C(t)$ in the case of qubits locally coupled to their environments
and $\eta_A(t) $=$\eta_B(t) $=$\eta_C(t) $ for non-local interaction of three subsystems with a common source of noise).
The single-qubit time-evolution operator $U_L\left(\eta_L,t\right)$ can be written as:
\begin{eqnarray}
\lefteqn{U_L\left(\eta_L,t\right)=\exp{\left[-i\int_0^t{\mathcal{H}_L(t^{\prime})} dt^{\prime}\right]}} \nonumber \\
 & & =\exp{\left(-i \epsilon t\right)}\left( \begin {array}{cc}  \cos{\phi_L(t)} & i\sin{\phi_L(t)} \\   i\sin{\phi_L(t)} & \cos{\phi_L(t)}  \end {array} \right),
\end{eqnarray}
where  $\hbar$=1 and $\phi_L(t)$=-$\nu \int_0^t \eta_L(t^{\prime}) dt^{\prime}$ is the random phase picked up
during the time interval $[0,t]$. The  evaluation of  the  time-evolved density matrix of Eq.~(\ref{rhoev}) requires the estimate
of the averaged terms of the type  $\langle [\cos{ \phi(t)}]^m [\sin{ \phi(t)]^k} \rangle$.
These can be expressed in terms of  the average of the phase factor $\langle e^{i n \phi(t)} \rangle$,
whose the explicit form reads~\cite{Bergli}:
\begin{equation} \label{ddtdif}
\langle e^{i n \phi(t)} \rangle=\langle \cos{n \phi(t)} \rangle +i\langle \sin{ n \phi(t)} \rangle
 \end{equation}
 with
 \begin{widetext}
 \begin{eqnarray} \label{ddtdif2}
 \langle \cos{n \phi(t)} \rangle&=&
G_{n}(t)=\left\{ \begin{array}{ccc}e^{-\gamma t}\left[\cosh {\left(\delta_{n\nu}t\right)} +\frac{\gamma}{\delta_{n\nu}}\sinh {\left(\delta_{n\nu}t\right)}\right] &\quad \textrm{for} \quad& \gamma > n\nu  \\  e^{-\gamma t}\left[\cos {\left(\delta_{n\nu}t\right)} +\frac{\gamma}{\delta_{n\nu}}\sin {\left(\delta_{n\nu}t\right)}\right] &\quad \textrm{for} \quad &\gamma < n\nu
\end{array}\right .  \nonumber\\
\langle \sin{ n \phi(t)} \rangle&=&0,
  \end{eqnarray}
  \end{widetext}
where $\delta_{n\nu}$=$\sqrt{|\gamma^2 -(n\nu)^2|}$  (with $m,k,n\in \mathbb{N}$).  Here, 
we examine the time-evolution of two classes of entangled states: the GHZ and  W Werner-type states.
The former reads:
\begin{equation} \label{grho}
 \rho_G(0)=r\rho_{GHZ}+\frac{1-r}{8} I_8,
 \end{equation}
 where  $\rho_{GHZ}$=$\frac{1}{2}\left(|000\rangle +|111\rangle  \right) \left( \langle 000 |+\langle 111 |\right)$,
  $r$ is the purity ranging from 0 to 1, and $I_8$ is the identity matrix of dimension 8.
 The explicit evaluation of the time-evolved state for both local and non-local system-environment coupling
 is reported in Appendix~\ref{app}. The different features  of the  classical environmental noise of three qubits
 plays a key role into the dynamics of GHZ-type state. When the three qubits are coupled to different environments,
 the  time-evolved density matrices  takes an X shape as shown in Eq.~\eqref{X1}. Instead when the interaction
 between the tripartite system and a common source of RTN is considered, all the elements
 of the three-qubit density matrix are non-vanishing after the initial time (see Eq.~\eqref{X2}). Such a behavior
 differs from what  was found in the two-qubit model~\cite{BenedettiAr}
 where the time-evolved density matrices of  any initial  Werner state has an X shape
regardless of  local and non-local qubit-environment interaction~\cite{Benedetti}.

In the other initial configuration, the
three qubits  are taken in the  W Werner-type state
\begin{equation} \label{wrho}
 \rho_W(0)=r\rho_{w}+\frac{1-r}{8} I_8,
 \end{equation}
 with  $\rho_{w}$=$\frac{1}{3}\left(|001\rangle +|010\rangle +|100\rangle  \right) \left( \langle 001| +\langle 010 |+\langle 100 |\right)$.
Unlike the GHZ Werner-type state, the time-evolved  density matrix of the system, reported in Eqs.~(\ref{rhoW1}) and (\ref{rhoW2})  
 takes the same form for local and non-local system-environment interaction.
  As it will be shown in the following,
the peculiar dynamics exhibited by  GHZ and  W Werner-type states 
is crucial in suppressing or preserving the quantum correlations among the three qubits.

It is worth noting that given the initial states of Eqs.~(\ref{grho}) and (\ref{wrho})
and the form of the  Hamiltonian written in Eq.~(\ref{mod1}),
the time-evolved states  are still invariant under the permutations of the three parties, 
regardless the local or non local nature of the system-environment coupling.
This allows us to evaluate more simply both tripartite negativity and quantum discord.

\section{Results}\label{results}

In this section, we evaluate the time evolution of the tripartite entanglement 
and quantum discord in the physical model  introduced in the previous
section.  The dynamics of quantum correlations is investigated
for the  GHZ- and W-type states
in the cases of local and non-local system-environment interaction.

\subsection{GHZ-type states}\subsubsection{Different Environments}
Here the effects of the dephasing on the state of Eq.~\eqref{grho} are examined.
When the qubits are coupled to different sources of RTN, the time-evolved
state takes an X shape which allows one to calculate easily the tripartite negative
and the expectation value of  the EW given in Eq.~\eqref{EW1}. Their analytical
forms are:
\begin{eqnarray}
& &\mathcal{N}^{(3)}(t)=\frac{1}{4} \max{\left[ 0,4rG^2_2(t)-(1-r)\right]} \nonumber \\
& &\textrm{Tr}[ \mathcal{\rho}_G(t)\mathcal{W}_{W2}]= -\frac{1}{4} \left[ 3rG^2_2(t)-\frac{3-r}{2}\right],
\end{eqnarray}
respectively. Such expressions  indicate that both  the tripartite negativity  and  
the expectation value of the $ \mathcal{W}_{W2}(t)$ can be expressed
in terms of $G^2_2(t)$. Specifically, we note that, at initial time  for $r$
ranging  from 1/5 to  3/7 the negativity   is different
from zero  while the expectation value of $\mathcal{W}_{W2}$ is zero or positive.  This means that the EW
does not detect the tripartite entanglement which is known to be present via the tripartite negativity measure.
This is in agreement with previous results in literature~\cite{Weinstein2010}.
In the limit of long times, $G^2_2(t)$ goes to zero, and, as a consequence $\mathcal{N}^{(3)}$  vanishes
and $-\langle\mathcal{W}_{W2}(t)\rangle$ becomes positive. In other words, the initial quantum correlations
in the form of entanglement disappear in the limit of long times.


\begin{figure}[htpb]
   \begin{center}
     \includegraphics[width=\linewidth]{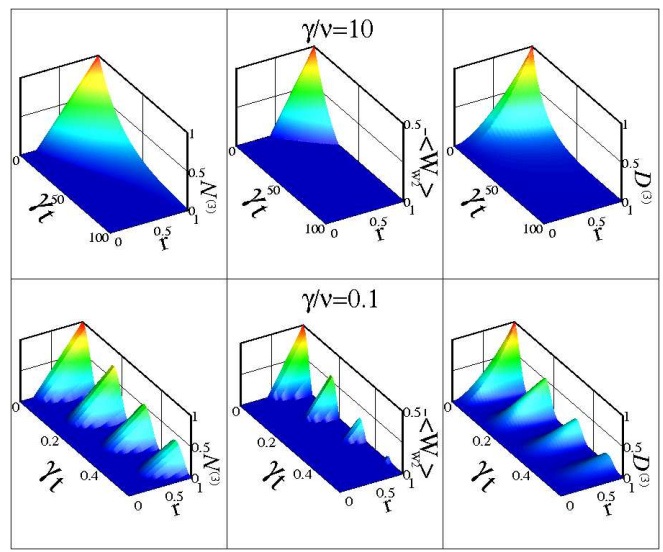}
           \caption{\label{fig1} (Color online) Top panels: 
    Negativity (left),  opposite of the expectation value of the EW $\mathcal{W}_{W2}$ (center), and  tripartite quantum discord (right)
    as a function of $\gamma t$ and the purity $r$ in the Markov regime  with $\gamma/\nu$=10,  when the three qubits,  initially prepared in the state $\rho_{G}(0)$, are     coupled to different sources of classical noise.
    Bottom panels:   same as in the top panels in the non-Markov  regime 
    with $\gamma/\nu=0.1$. }
  \end{center}
\end{figure}

 In Fig.~\ref{fig1}, we report $\mathcal{N}^{(3)}$, $-\langle \mathcal{W}_{W2}\rangle$, and $\mathcal{D}^{(3)}$
 as a function of the time and the initial purity of the state for two different values of $\gamma/\nu$, namely
 10 and 0.1, corresponding to the Markovian and non-Markovian dynamics of quantum correlations, respectively.
 In the first case, the RTN leads to a monotonic decay of entanglement and discord, while in the non-Markovian
regime all quantities are  damped oscillating functions of time and  display sudden death and revival phenomena.
Now let us compare qualitatively the behavior of the three estimators of quantum correlations. Even if the
tripartite negativity  and  quantum discord present  substantially the same qualitative behavior, nevertheless 
we  find that $\mathcal{D}^{(3)}$ can take non-vanishing values in regions whereas $\mathcal{N}^{(3)}$ is zero.  As it can clearly be seen
in the non-Markovian regime,   the peaks  of the tripartite negative found at a given value of $\gamma t$ decrease 
for smaller $r$'s  at higher rate than the corresponding peaks of the tripartite discord, and, as consequence vanish 
at smaller values of the purity of initial state. Such a behavior results to be consistent with was found  in bipartite systems
of initially entangled qubits  evolving under local decohering channels where separable states can exhibit  nonzero discord~\cite{Mazzola,Werlang,Fanchini2010}.
On the other hand, the entanglement and quantum correlations quantified in terms of $\mathcal{N}^{(3)}$ and $\mathcal{D}^{(3)}$
result to be higher, both in Markovian and non-Markovian regime, than the one detected by the  tripartite entanglement witnesses.

\subsubsection{Common Environment}
Now let us evaluate the dynamics of the quantum correlations
initially present  in the state $\rho_{G}(0)$ when the three qubits
are coupled to a common source of RTN. The time-evolved  tripartite negativity and  expectation value
of the witness $\mathcal{W}_{W2}$  can be expressed as:
\begin{eqnarray} \label{cocco}
& &\mathcal{N}^{(3)}(t)=\frac{1}{4} \max{\left[ 0,2r\sqrt{2\left(G^2_4(t)+1\right)}-(1-r)\right]} \nonumber \\
& & \textrm{Tr}[{\rho}_G(t)\mathcal{W}_{W2}]=- \frac{1}{2} \left[ \frac{3}{4}rG_4(t)-\left(\frac{3}{4}-r\right)\right].
\end{eqnarray}
It is worth noting that, unlike the  case of local system-environment interaction, here quantum correlations 
are not totally destroyed at long times. Indeed, the asymptotic forms of Eqs.~\eqref{cocco}, holding
for any degree of  qubit-environment coupling, are given by
\begin{eqnarray} \label{asy}
& &\lim_{t \rightarrow \infty}\mathcal{N}^{(3)}(t)=\frac{1}{4} \max{\left[ 0,(2\sqrt{2}+1)r-1\right]} \nonumber \\
& &-\lim_{t \rightarrow \infty} \textrm{Tr}[{\rho}_G(t)\mathcal{W}_{W2}]=\left(\frac{r}{2}-\frac{3}{8}\right).
\end{eqnarray}
This means that the saturation values of 
$\mathcal{N}^{(3)}$ and $\langle{\rho}_G\mathcal{W}_{W2}\rangle$
do not depend upon the parameter $\gamma$ that is, in terms, upon the Markovianity of the regimes.
 For an initial state with purity $r$ greater than $3/4$, the entanglement quantified in terms of the negativity 
 and that part of entanglement detectable by means of the witness $\mathcal{W}_{W2}$  do not vanish in the long time limit.
The partial preservation of the entanglement can be ascribed to the indirect  interaction among the qubits stemming
from the coupling  of the global system to a common noisy environment.  Unlike the local system-environment interaction,
here the environment is no longer only the source of decohering effects, but also 
represents a sort of interaction mediator between the subsystem. Such an interaction somehow 
hinders the destruction  of quantum correlations.  The survival of quantum correlations has  already been observed for
the bipartite entanglement of  two-qubit systems interacting with quantum environments~\cite{Maniscalco,Bellomo2008,Ma2012}.
When the purity ranges from $3/4$ to $1/(2\sqrt{2}+1)$,  the residual amount of  tripartite entanglement at large times
as quantified  by $\mathcal{N}^{(3)}$  is not detectable by means of the  EW $\mathcal{W}_{W2}$.


\begin{figure}[h]
   \begin{center}
     \includegraphics[width=\linewidth]{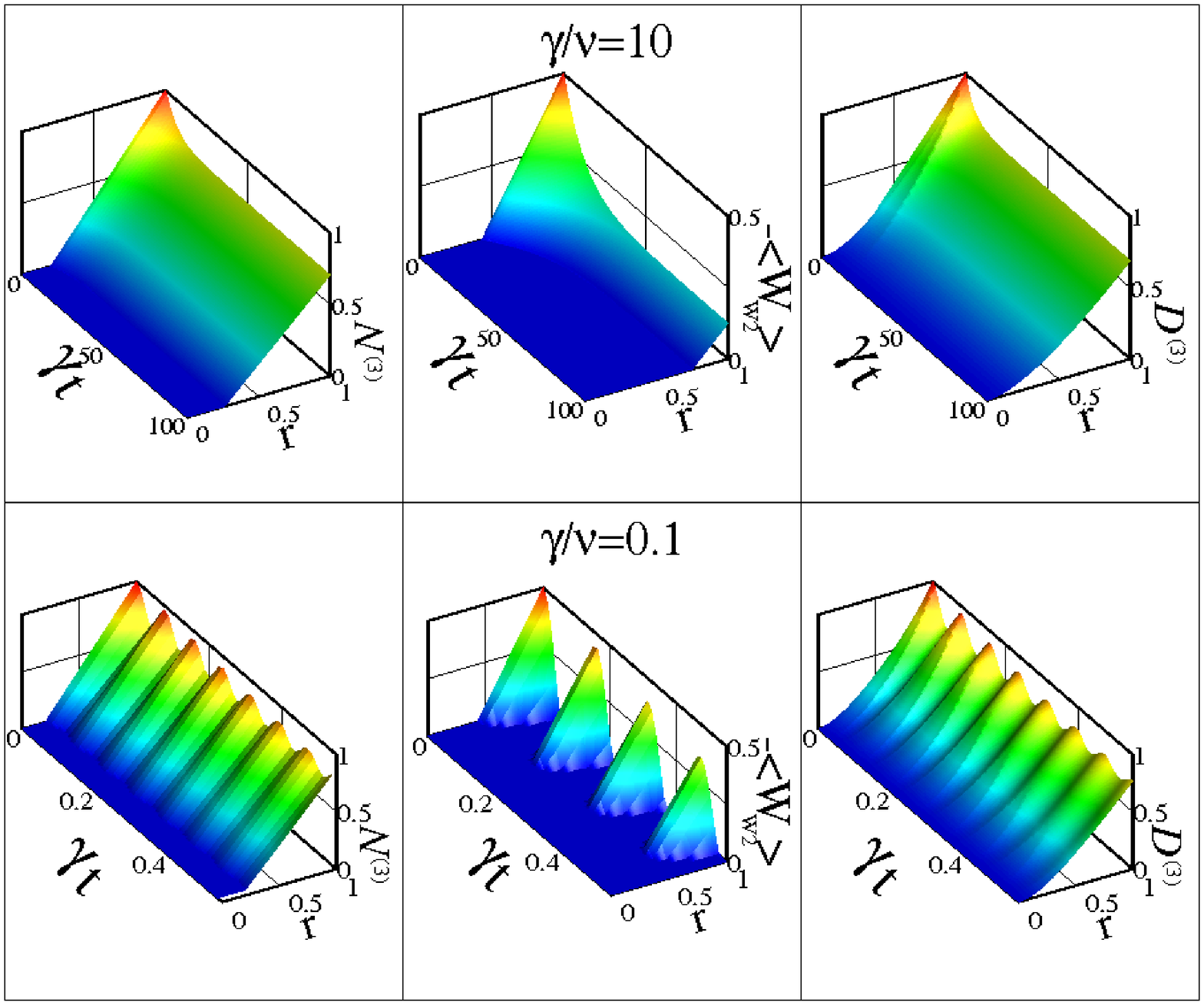}
           \caption{\label{fig2} (Color online) Top panels: 
    Negativity (left),  opposite of the expectation value of EW  $\mathcal{W}_{W2}$ (center), and  tripartite quantum discord (right)
    as a function of $\gamma t$ and the purity $r$ in the Markov regime  with $\gamma/\nu$=10,  when the three qubits,  initially prepared in the state $\rho_{GHZ}$, are coupled to a common source of classical noise.
     Bottom panels:  same as in the top panels in the non-Markov  regime 
    with $\gamma/\nu=0.1$.}
  \end{center}
\end{figure}

As shown in Fig.~\ref{fig2},  also  the tripartite quantum discord
can survive  the decohering effects due to the RTN.
In the Markovian regime, we find that all the estimators decay monotonically with time
until they reach the corresponding saturation values depending upon $r$.
In the non-Markovian  dynamics, both $\mathcal{N}^{(3)}$ and $\mathcal{D}^{(3)}$
exhibit  damped oscillations but  no sudden death of the correlations occurs
for not-separable initial states. Indeed,  the indirect interaction among the qubits
not only counteracts the total suppression of  of negativity and discord at long times but also prevents
 the disappearance of quantum correlations at finite times.  
%
The survival of $\mathcal{N}^{(3)}$ and $\mathcal{D}^{(3)}$ in the long time limit represents the major
  discrepancy with what was found in the two-qubit  form of the model of  Eqs.~(\ref{mod1}) and (\ref{mod2})~\cite{BenedettiAr}. There both  bipartite entanglement and discord disappear at long times,
 regardless the local or non-local character of  the interaction among the qubits and the environment.
 
 Finally,  the time behavior exhibited by the \emph{detectable} entanglement  in the non-Markovian regime
 appears to be peculiar. Indeed, the expectation value of the EW $\mathcal{W}_{W2}$  takes positive or null values
 at finite times in correspondence of the non-vanishing minima of the tripartite negativity.
 This implies that the suppression of  the sudden death of entanglement cannot be detected while,
 as indicated by Eqs.\eqref{asy}, the long-time entanglement protection can successfully
 be revealed.

 \subsection{W-type states}

\subsubsection{Different Environments}
The dynamics of the quantum correlations, initially present in the state $\rho_{W}(0)$
of Eq.~\eqref{wrho}, is here analyzed  for the case of qubit locally interacting with their
environments.  Unlike GHZ-type states discussed in the previous section, 
the tripartite negativity can not be put  in  a compact analytical form, while
the expectation value of the EW $\mathcal{W}_{W1}$ can be expressed as:
\begin{eqnarray}
\lefteqn{ \textrm{Tr}[{\rho}_W(t)\mathcal{W}_{W1}]=} \nonumber \\ & & -\frac{1}{24} \left[r \left(7G^3_2(t)+5G^2_2(t)+5 G_2(t)+4\right)-13\right].
\end{eqnarray}
At the initial time for $r$=1, the expectation value of $\mathcal{W}_{W1}$ is $-1/3$, and the tripartite negativity
is equal to 0.94. The EW detects entanglement when $r>13/21$, while $\mathcal{N}^{(3)}$ is nonzero for
$r>0.2096$.
\begin{figure}[htpb]
   \begin{center}
     \includegraphics*[width=\linewidth]{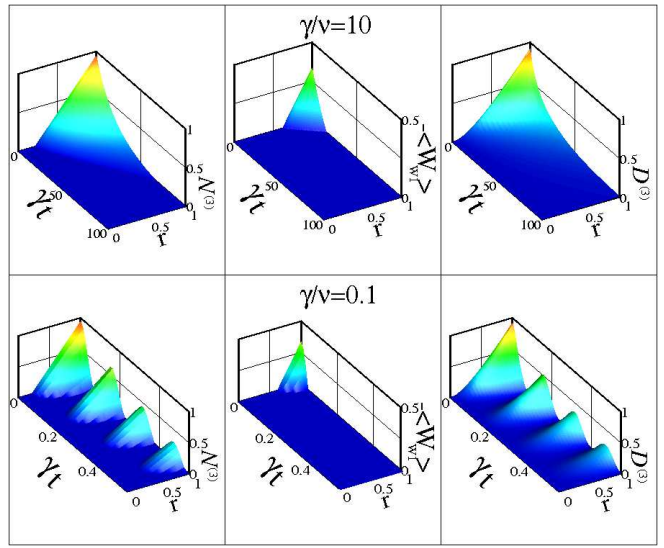}
           \caption{\label{fig3} (Color online) Top panels: 
    Negativity (left),  opposite of the expectation value of EW  $\mathcal{W}_{W1}$ (center),  and  tripartite quantum discord (right)
    as a function of $\gamma t$ and the purity $r$ in the Markov regime  with $\gamma/\nu$=10,  when the three qubits,  initially prepared in the state $\rho_{W}(0)$, are     coupled to different sources of classical noise.
     Bottom panels:  same as in the top panels in the non-Markovian  regime 
    with $\gamma/\nu=0.1$. }
  \end{center}
\end{figure}

Fig.~\ref{fig3}  displays  the tripartite negativity, the \emph{detectable} entanglement and the discord
 as a function of  time and  initial purity of the state for two different values of $\gamma/\nu$,
 corresponding to Markov and non-Markov regime. In the first case, we find again the monotonic
 decay of $\mathcal{N}^{(3)}$  and $\mathcal{D}^{(3)}$ both vanishing at very long times. On the
 other hand the witness  $\mathcal{W}_{W1}$ turns out to be unable to detect entanglement
 after a finite time. In the non-Markovian dynamics both the tripartite negativity and quantum
 discord exhibit revivals and sudden death phenomena, while the entanglement detectable
 by means of $\mathcal{W}_{W1}$  still decays monotonically with time and vanishes
 at short times. Also for this qubits configuration,  the  state can be separable even
 with a non-zero discord.
\subsubsection{Common Environment}
Finally, we  estimate the time evolution of  the quantum correlations 
when the three  qubits, initially set in the state $\rho_{W}(0)$, are subject
to the decohering effects of a   common source of RTN.  The expectation
value of the EW $\mathcal{W}_{W1}$ takes the form
\begin{eqnarray}
\lefteqn{ \textrm{Tr}[{\rho}_W(t)\mathcal{W}_{W1}]=} \nonumber \\ & & -\left[\frac{r}{32}  \left(9G_6(t)+6G_4(t)+7 G_2(t)+6\right)-\frac{13}{24}\right].
\end{eqnarray}
Unlike what was found in the previous section, at long times entanglement
cannot be detected independently from the value of $r$. 
\begin{figure}[htpb]
   \begin{center}
     \includegraphics*[width=\linewidth]{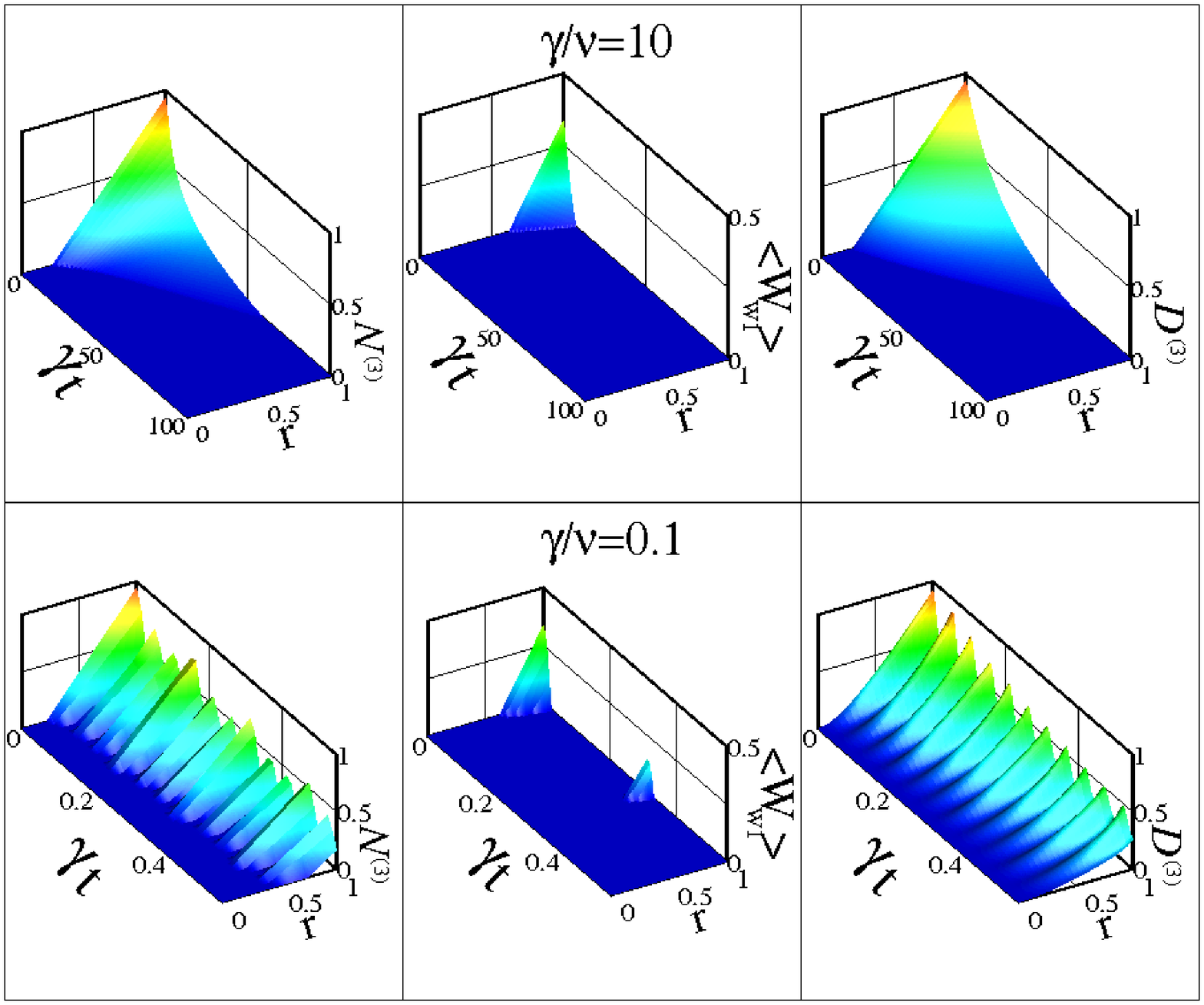}
           \caption{\label{fig4} (Color online) Top panels:  Negativity (left),  opposite of the expectation value of EW  $\mathcal{W}_{W1}$ (center), and  tripartite quantum discord (right)
    as a function of $\gamma t$ and the purity $r$ in the Markov regime  with $\gamma/\nu$=10,  when the three qubits,  initially prepared in the state $\rho_{W}$, are coupled to a common source of classical noise.
    Bottom panels: same as in the top panels in the non-Markovian  regime 
    with $\gamma/\nu=0.1$.  }
  \end{center}
\end{figure}
As it can be seen from Fig.~\ref{fig4},
in the  Markovian dynamics,  the usual decay of quantum correlations is observed
with  tripartite negativity and quantum discord vanishing at long times. When the non
Markovian dynamics is examined, $\mathcal{N}^{(3)}$ and $\mathcal{D}^{(3)}$ 
show damped oscillations without sudden death phenomena,
even if they vanish in the long time limit (not displayed in the bottom panels  of Fig.~\ref{fig4}).
 On the other hand,  a revival and sudden death of  the \emph{detectable} entanglement  are observed.
These results together with the ones shown in the previous subsection clearly
indicate that the quantum correlations initially present in the W-type states
are less robust than those of the GHZ-type states. Indeed, both
tripartite entanglement and quantum discord of the state  $\rho_{W}(0)$
are completely suppressed  not only in the case of subsystems  each locally coupled
to its environment but also for non-local system-environment coupling.

\section{Conclusions} \label{conclusions}

Recently, a number of theoretical works examined 
the dynamics of non-classical correlations in
tripartite systems coupled to external environments characterized by
different memory properties~~\cite{Casagrande, Altintas, Zhong-Xiao, Anza, An, Siomau,  Siomau2, Ma, Ann,Liu,Weinstein2009,Weinstein2010,Grimsmo}.
While different approaches have been adopted to
estimate entanglement~\cite{Casagrande, *Altintas, *Zhong-Xiao, *Anza, An, Siomau,*Siomau2,Weinstein2009,Weinstein2010}
  and Bell non-locality~\cite{Ma, Ann,Liu}, few studies focus on
the  time evolution of tripartite quantum correlations in terms of quantum discord~\cite{Grimsmo}.

In this paper, we address the dynamics
of quantum correlations in a model consisting of three initially entangled
qubits, not interacting among each other and  coupled to different  sources or to a common source
of  classical RTN.  Specifically, two initial configurations 
have been considered, namely  GHZ and W Werner-type states.
Unlike previous analyses~\cite{Siomau,Ann,Altintas,Liu,Ma,Weinstein2009}, here we have provided
an estimate of the tripartite quantum discord 
by using the approach recently introduced in Ref.~\cite{Giorgi}
which allows one to quantify the amount of  \emph{genuine} tripartite quantum correlations
of the system. These have  then been  compared with the entanglement
evaluated by means of the tripartite negativity  and with  the
detection ability of suitable EWs.

Our results show that both entanglement
and quantum discord are strongly affected
not only by the initial configuration of the qubits
but also by the local or non-local nature  of 
the system-environment interaction. In particular,
for a  GHZ-type initial state  
the indirect interaction between the qubits due to their
coupling to a common source of RTN allows for a long-time
tripartite entanglement  and quantum discord preservation, regardless
the Markov or non-Markov character of the environment itself. Furthermore,
we find that the survived entanglement can be detected by means
of  the witnesses. On the other hand,  quantum correlations
are completely destroyed   when  the subsystems are coupled to different environments.
The survival of entanglement and quantum discord and the disappearance
of sudden death phenomena, which have already
been observed in a number of bipartite systems interacting with quantum environments~\cite{Maniscalco,Bellomo2008,Ma2012}
are here closely related  to the tripartite nature of the system. Indeed,
in the two-qubit model analogous to the one here investigated
the number of quantum correlations vanish in the long time limit under any
condition~\cite{BenedettiAr}. 
By using approaches already used elsewhere~\cite{Zhou2011, Falci, Kuopanportti, Bergli}, it could certainly   be of interest to 
extend our investigation to the  case of $1/f^{\alpha}$ noises
stemming from a collection of random telegraph sources with
different switching rates. So, we could verify the survival of quantum correlations
when a large number  of decoherence channels is  considered.

%

In the other initial configuration, 
that is the qubits prepared in a W Werner-type state,
no preservation of quantum correlations is found.
Specifically, some of the  standard features of Markov and non-Markov
regime are observed. Indeed, the former exhibits a monotonic
decay of quantum correlation while in  the latter  sudden
death and revival  phenomena occur only when the  qubits  are locally coupled to
different sources of RTN. Anyway, both the
entanglement and  quantum discord of  W Werner-type state
turn out to be less robust than those of the GHZ-type state.

Finally,  the tripartite quantum discord
deserves a brief comment. It almost shows the same qualitative behavior
of the negativity for all the physical conditions examined. Nevertheless,
in agreement  with what was previously found in bipartite systems~\cite{Mazzola,Werlang,Fanchini2010,Ali}, 
the amount of tripartite quantum correlations quantified by discord
can be different from zero for  states without  entanglement.
From this point of view, our analysis seems to  further validate
the measure introduced in Ref.~\cite{Giorgi} as a good estimator
of the  \emph{genuine} tripartite quantum correlations which are not
entanglement.

\acknowledgments
The authors would like to thank  Claudia Benedetti, Matteo G.A. Paris, and Andrea Beggi for useful
discussions.

\appendix
\section{Evaluation of the time-evolved states} \label{app}
Here, we give the explicit forms of the time-evolved state
of the two different initial configurations of the qubits, namely GHZ and W Werner-type state,
for the case  of local and non-local system-environment interaction.

In order to evaluate the dynamics of the system,  
first we calculate the evolution of the initial state
for a  given choice of the noise parameter, as indicated
in  Eq.~(\ref{rhoev}). Then, the obtained density matrix  is averaged over noise
(see Eq.\eqref{ddtdif}).  For the input GHZ-type state  $\rho_G(0)$ of Eq.~\eqref{grho},
we find that, in the case of local system-environment coupling, the time-evolved density matrix of the system
takes the form
\begin{equation} \label{X1}
 \rho_G^{L}(t)= \left[ \begin {array}{cccccccc} \alpha(t)&0&0&0&0&0&0&\sigma(t)\\ \noalign{\medskip}0&\theta(t)&0&0&0&0&\beta(t)&0\\ \noalign{\medskip}0&0&\theta(t)&0&0&\beta(t)&0&0\\ \noalign{\medskip}0&0&0&\theta(t)&\beta(t)&0&0&0\\ \noalign{\medskip}0&0&0&\beta(t) &\theta(t)&0&0&0
\\ \noalign{\medskip}0&0& \beta(t)&0&0&\theta(t)&0&0\\ \noalign{\medskip}0&\beta(t) &0&0&0&0&\theta(t)&0\\ \noalign{\medskip}\sigma(t) &0&0&0&0&0&0&\alpha(t)\end {array} \right] ,
\end{equation}
 with
   \begin{eqnarray}
\lefteqn{ \alpha(t)=\frac{1}{8}\left(1+ 3 r G_{2}^2(t)\right),  \quad \beta(t)=\frac{r}{8}\left(1- G_{2}^2(t)\right),} \nonumber \\
{ } & & \theta(t)=\frac{1}{8}\left(1- rG_{2}^2(t)\right),  \quad \textrm{and} \quad\sigma(t) =\frac{r}{8}\left(1+ 3  G^2_{2}(t)\right). \nonumber \\
 \end{eqnarray}

On the other hand, for the case of non local qubit-environment interaction
the dynamics of the system  initially in $\rho_G(0)$ results into:
\begin{widetext}
 \begin{equation} \label{X2}
 \rho_G^{NL}(t)= \left[ \begin {array}{cccccccc}\frac{1}{8} +3\mu (t)&-\lambda(t)&-\lambda(t)&-\lambda(t)&-\lambda(t)&-\lambda(t)&-\lambda(t)&\frac{r}{8} +3\mu (t)\\ \noalign{\medskip}-\lambda(t)&\frac{1}{8} -\mu (t)&\lambda(t)&\lambda(t)&\lambda(t)&\lambda(t)&\lambda(t)&-\lambda(t)\\ \noalign{\medskip}-\lambda(t)&\lambda(t)&\frac{1}{8} -\mu (t)&\lambda(t)&\lambda(t)&\lambda(t)&\lambda(t)&-\lambda(t)\\ \noalign{\medskip}-\lambda(t)&\lambda(t)&\lambda(t)&\frac{1}{8} -\mu (t)&\lambda(t)&\lambda(t)&\lambda(t)&-\lambda(t)\\ \noalign{\medskip}-\lambda(t)&\lambda(t)&\lambda(t)&\lambda(t) &\frac{1}{8} -\mu (t)&\lambda(t)&\lambda(t)&-\lambda(t)
\\ \noalign{\medskip}-\lambda(t)&\lambda(t)& \lambda(t)&\lambda(t)&\lambda(t)&\frac{1}{8} -\mu (t)&\lambda(t)&-\lambda(t)\\ \noalign{\medskip}-\lambda(t)&\lambda(t)&\lambda(t)&\lambda(t)&\lambda(t)&\lambda(t)&\frac{1}{8} -\mu (t)&-\lambda(t)\\ \noalign{\medskip}\frac{r}{8} +3\mu (t) &-\lambda(t)&-\lambda(t)&-\lambda(t)&-\lambda(t)&-\lambda(t)&-\lambda(t)&\frac{1}{8} +3\mu (t)\end {array} \right] ,
\end{equation}
\end{widetext}
where
\begin{equation}
 \mu (t)=\frac{r}{16}\left(1+G_{4}(t)\right), \quad \textrm{and} \quad \lambda (t)=\frac{r}{16}\left(1-G_{4}(t)\right).
 \end{equation}
 These results clearly show that the form of the time-evolved density matrix of three qubits initially
 in a GHZ-type state depend upon the local or non-local nature of the system-environment coupling.

 Now let us focus on the time evolution of the system initially
 prepared in the W-type state  $\rho_W(0)$ given in Eq.~\eqref{wrho}.
 When each qubit interacts locally with its environment, the density matrix of the system at time $t$  
takes the form:
\begin{widetext}
 \begin{equation} \label{rhoW1}
 \rho_W^{L}(t)= \left[ \begin {array}{cccccccc}\frac{\kappa(t)}{8}+\varsigma&0&0&\frac{\kappa(t)}{12}&0&\frac{\kappa(t)}{12}&\frac{\kappa(t)}{12}&0\\ \noalign{\medskip}0&\frac{\tau(t)}{24}+\varsigma&\frac{\iota(t)}{12}&0&\frac{\iota(t)}{12}&0&0&\frac{\chi(t)}{12}\\ \noalign{\medskip}0&\frac{\iota(t)}{12}&\frac{\tau(t)}{24}+\varsigma&0&\frac{\iota(t)}{12}&0&0&\frac{\chi(t)}{12}\\ \noalign{\medskip}\frac{\kappa(t)}{12}&0&0&\frac{\xi(t)}{24}+\varsigma&0&\frac{\varphi(t)}{12}&\frac{\varphi(t)}{12}&0\\ \noalign{\medskip}0&\frac{\iota(t)}{12}&\frac{\iota(t)}{12}&0 &\frac{\tau(t)}{24}+\varsigma&0&0&\frac{\chi(t)}{12}
\\ \noalign{\medskip}\frac{\kappa(t)}{12}&0&0&\frac{\varphi(t)}{12}&0&\frac{\xi(t)}{24}+\varsigma&\frac{\varphi(t)}{12}&0\\ \noalign{\medskip}\frac{\kappa(t)}{12}&0 &0&\frac{\varphi(t)}{12}&0&\frac{\varphi(t)}{12}&\frac{\xi(t)}{24}+\varsigma&0\\ \noalign{\medskip}0&\frac{\chi(t)}{12}&\frac{\chi(t)}{12}&0&\frac{\chi(t)}{12}&0&0&\frac{\chi(t)}{8}+\varsigma\end {array} \right] ,
\end{equation}
\end{widetext}
where
\begin{widetext}
\begin{eqnarray}
 & &\kappa(t)=r\big(1+G_{2}(t)\big)^2\big(1-G_{2}(t)\big), \quad \varsigma=\frac{1}{8}(1-r), \quad \chi(t)=r\big(1-G_{2}(t)\big)^2\big(1+G_{2}(t)\big), \nonumber \\
& & \tau(t)=r\big(1+G_{2}(t)\big)\big[\big(1+G_{2}(t)\big)^2+2\big(1-G_{2}(t)\big)^2 \big], \quad
 \iota(t)=r\big(1+G_{2}(t)\big)\left(1+G_{2}^2(t)\right), \nonumber \\ & & \
 \xi(t)=r\big(1-G_{2}(t)\big)\big[2\big(1+G_{2}(t)\big)^2+\big(1-G_{2}(t)\big)^2 \big], \quad  \varphi(t)=r\big(1-G_{2}(t)\big)\left(1+G_2^2(t)\right). \nonumber
  \end{eqnarray}
  \end{widetext}
When all the three qubits are coupled to a common source of RTN, the time-evolved state
of the system can written as:
\begin{widetext}
\begin{equation} \label{rhoW2}
 \rho_W^{NL}(t)= \left[ \begin {array}{cccccccc}\frac{1}{8}+\Lambda(t)&0&0&\Xi(t)&0&\Xi(t)&\Xi(t)&0\\ \noalign{\medskip}0&\frac{1}{8}+\Upsilon(t)&\Omega(t)&0&\Omega(t)&0&0&\Gamma(t)\\ \noalign{\medskip}0&\Omega(t)&\frac{1}{8}+\Upsilon(t)&0&\Omega(t)&0&0&\Gamma(t)\\ \noalign{\medskip}\Xi(t)&0&0&\frac{1}{8}+\Phi(t)&0&\Delta(t)&\Delta(t)&0\\ \noalign{\medskip}0&\Omega(t)&\Omega(t)&0 &\frac{1}{8}+\Upsilon(t)&0&0&\Gamma(t)
\\ \noalign{\medskip}\Xi(t)&0&0&\Delta(t)&0&\frac{1}{8}+\Phi(t)&\Delta(t)&0\\ \noalign{\medskip}\Xi(t)&0 &0&\Delta(t)&0&\Delta(t)&\frac{1}{8}+\Phi(t)&0\\ \noalign{\medskip}0&\Gamma(t)&\Gamma(t)&0&\Gamma(t)&0&0&\frac{1}{8}+\Psi(t)\end {array} \right] ,
\end{equation}
\end{widetext}
with
\begin{widetext}
\begin{eqnarray}
 & &\Lambda(t)=r\left(\frac{1}{16}+\frac{3G_{2}(t)}{32}-\frac{3G_{4}(t)}{16}-\frac{3G_{6}(t)}{32}\right), \quad \Xi(t)=r\left(\frac{1}{16}+\frac{3G_{2}(t)}{32}-\frac{G_{4}(t)}{16}-\frac{3G_{6}(t)}{32}\right), \nonumber \\
& & \Upsilon(t)=r\left(\frac{-1}{48}+\frac{7G_{2}(t)}{96}+\frac{G_{4}(t)}{16}+\frac{3G_{6}(t)}{32}\right), \quad
 \Omega(t)=r\left(\frac{5}{48}+\frac{7G_{2}(t)}{96}+\frac{G_{4}(t)}{16}+\frac{3G_{6}(t)}{32}\right), \nonumber \\
  & & \Gamma(t)=r\left(\frac{1}{16}-\frac{3G_{2}(t)}{32}-\frac{G_{4}(t)}{16}+\frac{3G_{6}(t)}{32}\right), \quad
\Phi(t)=r\left(\frac{-1}{48}-\frac{7G_{2}(t)}{96}+\frac{G_{4}(t)}{16}-\frac{3G_{6}(t)}{32}\right), \nonumber \\
& & \Delta(t)=r\left(\frac{5}{48}-\frac{7G_{2}(t)}{96}+\frac{G_{4}(t)}{16}-\frac{3G_{6}(t)}{32}\right),  \quad \Psi(t)=r\left(\frac{1}{16}-\frac{3G_{2}(t)}{32}-\frac{3G_{4}(t)}{16}+\frac{3G_{6}(t)}{32}\right).\nonumber 
\end{eqnarray} 
\end{widetext}
Unlike the GHZ Werner-type state, we note that in this case the time-evolved density matrix
of the system takes the same form for local and non-local system-environment interaction. Indeed, in both the matrices of  Eqs.~(\ref{rhoW1}) and (\ref{rhoW2})
 the  diagonal  4$\times$4 subblocks  have an X shape, while  in the antidiagonal ones
the diagonal and antidiagonal elements are zero.

 \end{document}